# Review of Artificial Intelligence Techniques in Imaging Data Acquisition, Segmentation and Diagnosis for COVID-19

Feng Shi [†], Jun Wang [†], Jun Shi [†], Ziyan Wu, Qian Wang, Zhenyu Tang, Kelei He, Yinghuan Shi, Dinggang Shen[*]

***Abstract*—The pandemic of coronavirus disease 2019 (COVID-19) is spreading all over the world. Medical imaging such as X-ray and computed tomography (CT) plays an essential role in the global fight against COVID-19, whereas the recently emerging artificial intelligence (AI) technologies further strengthen the power of the imaging tools and help medical specialists. We hereby review the rapid responses in the community of medical imaging (empowered by AI) toward COVID-19. For example, AI-empowered image acquisition can significantly help automate the scanning procedure and also reshape the workflow with minimal contact to patients, providing the best protection to the imaging technicians. Also, AI can improve work efficiency by accurate delination of infections in X-ray and CT images, facilitating subsequent quantification. Moreover, the computer-aided platforms help radiologists make clinical decisions, i.e., for disease diagnosis, tracking, and prognosis. In this review paper, we thus cover the entire pipeline of medical imaging and analysis techniques involved with COVID-19, including image acquisition, segmentation, diagnosis, and follow-up. We particularly focus on the integration of AI with X-ray and CT, both of which are widely used in the frontline hospitals, in order to depict the latest progress of medical imaging and radiology fighting against COVID-19.***

*Index Terms*—COVID-19, artificial intelligence, image acquisition, segmentation, diagnosis

This work was supported in part by the Shanghai Science and Technology Foundation (18010500600, 19QC1400600), National Key Research and Development Program of China (2018YFC0116400), and Natural Science Foundation of Jiangsu Province (BK20181339). ([*] Corresponding author: Dinggang Shen)

[†] F. Shi, J. Wang, and J. Shi contributed equally to this work.

F. Shi and D. Shen are with the Department of Research and Development, Shanghai United Imaging Intelligence Co., Ltd., Shanghai 200232, China (e-mail: feng.shi@united-imaging.com; Dinggang.Shen@gmail.com).

J. Wang and J. Shi are with Key Laboratory of Specialty Fiber Optics and Optical Access Networks, Shanghai Institute for Advanced Communication and Data Science, School of Communication and Information Engineering, Shanghai University, Shanghai 200444, China (e-mail: wangjun_shu@shu.edu.cn; junshi@shu.edu.cn).

Z. Wu is with United Imaging Intelligence, Cambridge, MA 02140, USA (e-mail: ziyan.wu@united-imaging.com).

Q. Wang is with the Institute for Medical Imaging Technology, School of Biomedical Engineering, Shanghai Jiao Tong University, Shanghai 200030, China (e-mail: wang.qian@sjtu.edu.cn).

Z. Tang is with Beijing Advanced Innovation Center for Big Data and Brain Computing, Beihang University, Beijing 100191, China (e-mail: tangzhenyu@buaa.edu.cn).

K. He is with the Medical School of Nanjing University, Nanjing, China. He is also with National Institute of Healthcare Data Science at Nanjing University, Nanjing 210093, China (e-mail: hkl@nju.edu.cn).

Y. Shi is with the National Key Laboratory for Novel Software and Technology, Nanjing University, Nanjing, China. He is also with the National Institute of Healthcare Data Science at Nanjing University, Nanjing 210093, China (e-mail: syh@nju.edu.cn).

## I. Introduction

THE coronavirus disease 2019 (COVID-19), caused by severe acute respiratory syndrome coronavirus 2 (SARS-CoV-2), is an ongoing pandemic. The number of people infected by the virus is increasing rapidly. Up to April 5, 2020, 1,133,758 cases of COVID-19 have been reported in over 200 countries and territories, resulting in approximately 62,784 deaths (with a fatal rate of 5.54%) [1]. This has led to great public health concern in the international community, as the World Health Organization (WHO) declared the outbreak to be a Public Health Emergency of International Concern (PHEIC) on January 30, 2020 and recognized it as a pandemic on March 11, 2020 [2, 3].

Reverse Transcription-Polymerase Chain Reaction (RT-PCR) test serves as the gold standard of confirming COVID-19 patients [4]. However, the RT-PCR assay tends to be inadequate in many areas that have been severely hit especially during early outbreaks of this disease. The lab test also suffers from high false-negative rates, due to many factors, such as sample preparation and quality control [5]. In clinical practice, easily accessible imaging equipments, such as chest X-ray and thoracic CT, provide huge assistance to clinicians [6-11]. Particularly in China, many cases were identified as suspected of COVID-19, if characteristic manifestations in CT scans were observed [5]. The suspected patients, even without clinical symptoms (e.g., fever and coughing), were also hospitalized or quarantined for further lab tests. Given the high false-positive rate of the nucleic acid tests, many suspected patients have to be tested multiple times several days apart before reaching a confident diagnosis. Hence, the imaging findings play a critical role in constraining the viral transmission and also fighting against COVID-19.

The workflow of imaging-based diagnosis for COVID-19, taking thoracic CT as an example, includes three stages in general, i.e., 1) pre-scan preparation, 2) image acquisition, and 3) disease diagnosis. In the pre-scan preparation stage, each subject is instructed and assisted by a technician to pose on the patient bed according to a given protocol. In the image acquisition stage, CT images are acquired during a single breath-hold. The scan ranges from the apex to the lung base. Scans are done from the level of the upper thoracic inlet to the inferior level of the costophrenic angle with the optimized parameters set by the radiologist(s), based on the patient's body shape. From the acquired raw data, CT images are reconstructed and then transmitted through picture archiving and communication systems (PACS) for subsequent reading and diagnosis.

Artificial intelligence (AI), an emerging technology in the field of medical imaging, has contributed actively to fight COVID-19 [12]. Compared to the traditional imaging workflow that heavily relies on the human labors, AI enables more safe, accurate and efficient imaging solutions. Recent AI-empowered applications in COVID-19 mainly include the dedicated imaging platform, the lung and infection region segmentation, the clinical assessment and diagnosis, as well as the pioneering basic and clinical research. Moreover, many commercial products have been developed, which successfully integrate AI to combat COVID-19 and clearly demonstrate the capability of the technology. The Medical Imaging Computing Seminar (MICS) [1], a China's leading alliance of medical imaging scholars and start-up companies, organized this first online seminar on COVID-19 on Febuary 18, 2020, which attracted more than ten thousands of visits. All the above examples show the tremendous enthusiasm cast by the public for AI-empowered progress in the medical imaging field, especially during the ongoing pandemic.

Due to the importance of AI in all the spectrum of the imaging-based analysis of COVID-19, this review aims to extensively discuss the role of medical imaging, especially empowered by AI, in fighting the COVID-19, which will inspire future practical applications and methodological research. In the following, we first introduce intelligent imaging platforms for COVID-19, and then summarize popular machine learning methods in the imaging workflow, including segmentation, diagnosis and prognosis. Several publicly available datasets are also introduced. Finally, we discuss several open problems and challenges. We expect to provide guidance for researchers and radiologists through this review. Note that we review the most related medical-imaging-based COVID-19 studies up to March 31, 2020.

## II. AI-empowered Contactless Imaging Workflows

Healthcare practitioners are particularly vulnerable concerning the high risk of occupational viral exposure. Imaging specialists and technicians are of high priority, such that any potential contact with the virus could be under control. In addition to the personal protective equipment (PPE), one may consider dedicated imaging facilities and workflows, which are significantly important to reduce the risks and save lives.

### A. Conventional Imaging Workflow

Chest X-ray and CT are widely used in the screening and diagnosis of COVID-19 [6-11]. It is important to employ a contactless and automated image acquisition workflow to avoid the severe risks of infection during COVID-19 pandemic. However, the conventional imaging workflow includes inevitable contact between technicians and patients. Especially, in patient positioning, technicians first assist in posing the patient according to a given protocol, such as head-first versus feet-first, and supine versus prone in CT, followed by visually identifying the target body part location on the patient and manually adjusting the relative position and pose between the patient and the X-ray tube. This process puts the technicians in close contact with the patients, which leads to high risks of viral exposure. Thus, a contactless and automated imaging workflow is needed to minimize the contact.

### B. AI-Empowered Imaging Workflow

Many modern X-ray and CT systems are equipped with cameras for patient monitoring purposes [13-16]. During the outbreak of COVID-19, those devices provided chance to establish a contactless scanning workflow. Technicians can monitor the patient from the control room via a live video stream from the camera. However, from only the overhead view of the camera, it is still challenging for the technician to determine the scanning parameters such as scan range. In this case, AI is able to automate the process [17-25] by identifying the pose and shape of the patient from the data acquired with visual sensors such as RGB, Time-of-Flight (TOF) pressure imaging [26] or thermal (FIR) cameras. Thus, the optimal scanning parameters can be determined.

One typical scanning parameter that can be estimated with AI-empowered visual sensors is the scan range that defines the starting and ending positions of the CT scan. Scan range can be identified by detecting anatomical joints of the subject from the images. Much recent work [27-29] has focused on estimating the 2D [30-35] or 3D keypoint locations [28, 36-39] on the patient body. These keypoint locations usually include major joints such as the neck, shoulders, elbows, ankles, wrists, and knees. Wang *et al.* [40] has shown that such an automated workflow can significantly improve scanning efficiency and reduce unnecessary radiation exposure. However, such keypoints usually represent only a very sparse sampling of the full 3D mesh [41] in the 3D space (that defines the digital human body).

Other important scanning parameters can be inferred by AI, including ISO-centering. ISO-centering refers to aligning the target body region of the subject, so that the center of the target body region overlaps with the scanner ISO center and thus the overall imaging quality is optimal. Studies have shown that, with better ISO-centering, radiation dosage can be reduced while maintaining similar imaging quality [42]. In order to align the target body region to the ISO center, and given that anatomical keypoints usually represent only a very sparse sampling of the full 3D mesh in the 3D space (defining the digital human body), Georgakis *et al.* [43] propose to recover human mesh from a single monocular RGB image using a parametric human model SMPL [44]. Unlike other related studies [45], they employ a hierarchical kinematic reasoning for each kinematic chain of the patient to iteratively refine the estimation of each anatomical keypoint to improve the system robustness to clutters and partial occlusions around the joints of the patient. Singh *et al.* [18] present a technique, using depth sensor data, to retrieve a full 3D patient mesh by fitting the depth data to a parametric human mesh model based on anatomical landmarks detected from RGB image. One recent solution proposed by Ren *et al.* [41] learns a model that can be trained just once and have the capability to be applied across multiple such applications based on dynamic multi-modal inference.

---

[1] http://www.mics.net.cn/

With this framework in application with an RGB-depth input sensor, even if one of the sensor modalities fails, the model above can still perform 3D patient body inference with the remaining data.

*C. Applications in COVID-19*

During the outbreak of COVID-19, several essential contactless imaging workflows were established[17, 40, 41], from the utilization of monitoring cameras in the scan room [13-15, 27], or on the device [46], to mobile CT platforms [17, 46-49] with better access to patients and flexible installation.

A notable example is an automated scanning workflow based on a mobile CT platform empowered by visual AI technologies [17], as shown in Fig. 1(a). The mobile platform is fully self-contained with an AI-based pre-scan and diagnosis system [46]. It was redesigned into a fully isolated scan room and control room. Each room has its own entrance to avoid any unnecessary interaction between technicians and patients.

After entering the scan room, the patient is instructed, by visual and audio prompts, to pose on the patient bed (Fig. 1(b)). Technicians can observe through the window and also the live video transmitted from the ceiling-mounted AI camera in the scan room, and correct the pose of the patient if necessary (Fig. 1(c)). Once the patient is deemed ready, either by the technician or the motion analysis algorithm, the patient positioning algorithm will automatically recover the 3D pose and fully-reconstructed mesh of the patient from the images captured with the camera [41]. Based on the 3D mesh, both the scan range and the 3D centerline of the target body part of the patient are estimated and converted into control signals and optimized scanning parameters for the technician to verify. If necessary, the technician can make adjustments. Once verified, the patient bed will be automatically aligned to ISO center and moved into CT gantry for scanning. After CT images are acquired, they will be processed and analyzed for screening and diagnosis purposes.

### III. AI IN IMAGE SEGMENTATION

Segmentation is an essential step in AI-based COVID-19 image processing and analysis. It delineates the regions of interest (ROIs), e.g., lung, lobes, bronchopulmonary segments, and infected regions or lesions, in the chest X-ray or CT images for further assessment and quantification.

CT provides high-quality 3D images for detecting COVID-19. To segment ROIs in CT, deep learning methods are widely used. The popular segmentation networks for COVID-19 include classic U-Net [50-55], UNet++ [56, 57], VB-Net [58]. Compared with CT, X-ray is more easily accessible around the world. However, due to the ribs projected onto soft tissues in 2D and thus confounding image contrast, the segmentation of X-ray images is even more challenging. Currently, there is no method developed for segmenging X-ray images for COVID-19. However, Gaal *et al.* [59] adopt an Attention-U-Net for lung segmentation in X-ray images for pneumonia, and although the research is not specified for COVID-19, the method can be applied to the diagnosis of COVID-19 and other diseases easily.

Although now there are limited segmentation works directly related to COVID-19 , many papers consider segmentation as a necessary process in analyzing COVID-19. Table I summarizes representative works involving image segmentation in COVID-19 studies.

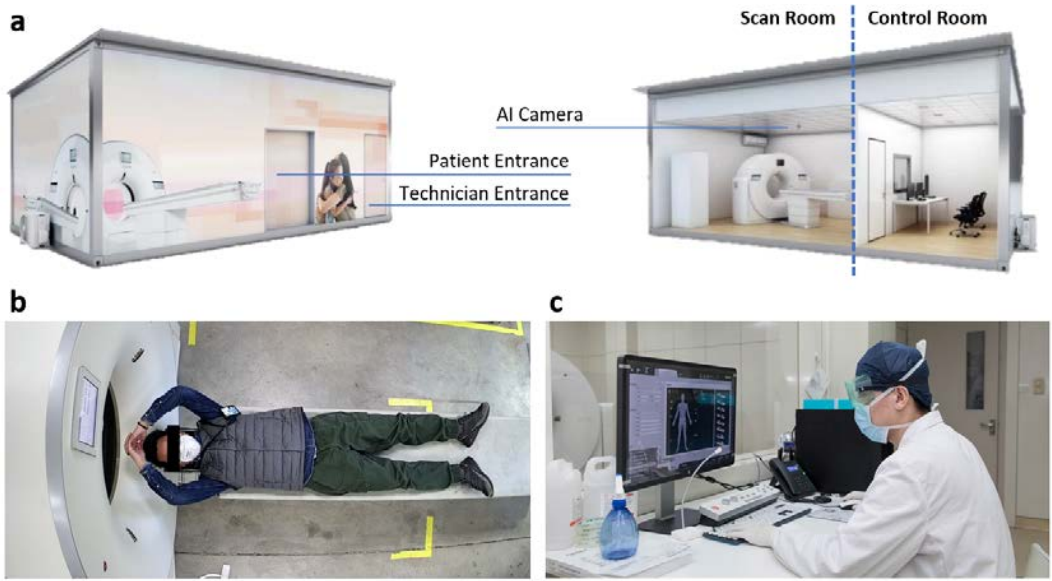

Fig. 1. (a) A mobile CT platform equipped with AI-empowered automated image acquisition workflow; (b) An example image captured by patient monitoring camera of CT system; (c) Positioning and scanning of patient operated remotely by a technician.

TABLE I
SUMMARY OF IMAGE SEGMENTATION METHODS IN COVID-19 APPLICATIONS

| Literature | Modality | Method | Target ROI | Application | Highlights |
|---|---|---|---|---|---|
| Zheng *et al.* [50] | CT | U-Net | Lung | Diagnosis | Weakly-supervised method by pseudo labels |
| Cao *et al.* [51] | CT | U-Net | Lung / Lesion | Quantification | |
| Huang *et al.* [52] | CT | U-Net | Lung / Lung lobes / lesion | Quantification | |
| Qi *et al.* [53] | CT | U-Net | Lung lobes / Lesion | Quantification | |
| Gozes *et al.* [54] | CT | U-Net/ Commercial Software | Lung / Leson | Diagnosis | Combination of 2D and 3D methods |
| Li *et al.* [55] | CT | U-Net | Lesion | Diagnosis | |
| Chen *et al.* [56] | CT | UNet++ | Lesion | Diagnosis | |
| Jin *et al.* [57] | CT | UNet++ | Lung / Lesion | Diagnosis | Joint segmentation and classification |
| Shan *et al.* [58] | CT | VB-Net | Lung / Lung lobes / Lung segments / Lesion | Quantification | Human-in-the-loop |
| Tang *et al.* [60] | CT | Commercial Software | Lung / Lesion / Trachea / Bronchus | Quantification | |
| Shen *et al.* [61] | CT | Threshold-based region growing [62] | Lesion | Quantification | |

*A. Segmentation of Lung Regions and Lesions*

In terms of target ROIs, the segmentation methods in COVID-19 applications can be mainly grouped into two categories, i.e., the lung-region-oriented methods and the lung-lesion-oriented methods. The lung-region-oriented methods aim to separate lung regions, i.e., whole lung and lung lobes, from other (background) regions in CT or X-ray, which is considered as a pre-requisite step in COVID-19 applications [50-54, 57, 58, 60]. For example, Jin *et al.* [57] propose a two-stage pipeline for screening COVID-19 in CT images, in which the whole lung region is first detected by an efficient segmentation network based on UNet++. The lung-lesion-oriented methods aim to separate lesions (or metal and motion artifacts) in the lung from lung regions [51-58, 60, 61]. Because the lesions or nodules could be small with a variety of shapes and textures, locating the regions of the lesions or nodules is required and has often been considered a challenging detection task. Notably, in addition to segmentation, the attention mechanism is reported as an efficient localization method in screening [59], which can be adopted in COVID-19 applications.

*B. Segmentation Methods*

In the literature, there have been numerous techniques for lung segmentation with different purposes [63-67]. The U-Net is a commonly used technique for segmenting both lung regions and lung lesions in COVID applications [50-53]. The U-Net, a type of fully convolutional network proposed by Ronneberger [68], has a U-shape architecture with symmetric encoding and decoding signal paths. The layers of the same level in two paths are connected by the shortcut connections. In this case, the network can therefore learn better visual semantics as well as detailed contextures, which is suitable for medical image segmentation.

Various U-Net and its variants have been developed, achieving reasonable segmentation results in COVID-19

applications. Çiçek *et al.* [63] propose the 3D U-Net that uses the inter-slice information by replacing the layers in conventional U-Net with a 3D version. Milletari *et al.* [64] propose the V-Net which utilizes the residual blocks as the basic convolutional block, and optimize the network by a Dice loss. By equipping the convolutional blocks with the so-called bottleneck blocks, Shan *et al.* [58] use a VB-Net for more efficient segmentation. Zhou *et al.* [65] propose the UNet++, which is much more complex than U-Net, as the network inserts a nested convolutional structure between the encoding and decoding path. Obviously, this type of network can improve the performance of segmentation. However, it is more difficult to train. This network is also used for locating lesions in COVID-19 diagnosis [56]. Recently advanced attention mechanisms can learn the most discriminant part of the features in the network. Oktay *et al.* [67] propose an Attention U-Net that is capable of capturing fine structures in medical images, thereby suitable for segmenting lesions and lung nodules in COVID-19 applications.

Training a robust segmentation network requires sufficient labeled data. In COVID-19 image segmentation, adequate training data for segmentation tasks is often unavailable since manual delineation for lesions is labor-intensive and time-consuming. To address this, a straightforward method is to incorporate human knowledge. For example, Shan *et al.* [58] integrate human-in-the-loop strategy into the training of a VB-net based segmentation network, which involves interactivity with radiologists into the training of the network. Qi *et al.* [53] delineate the lesions in the lung using U-Net with the initial seeds given by a radiologist. Several other works used diagnostic knowledge and identified the infection regions by the attention mechanism [57]. Weakly-supervised machine learning methods are also used when the training data are insufficient for segmentation. For example, Zheng *et al.* [50] propose to use an unsupervised method to generate pseudo segmentation masks for the images. As lacking of annotated medical images is common in lung segmentation, unsupervised and semi-supervised methods are highly demanded for COVID-19 studies.

*C. Applications in COVID-19*

Segmentation can be used in various COVID-19 applications, among which diagnosis is frequently reported [50, 54-57, 69, 70]. For example, Li *et al.* [55] use U-Net for lung segmentation in a multi-center study for distinguishing COVID-19 from community-acquired pneumonia on Chest CT. Jin *et al.* propose an AI system for fast COVID-19 diagnosis [57]. The input to the classification model is the CT slices that have been segmented by a segmentation network.

Another application of image segmentation is quantification [51-53, 58, 60, 61], which further serves for many medical applications. For example, Shan *et al.* [58] propose a VB-Net for segmentation of lung, lung lobes and lung infection, which provide accurate quantification data for medical studies, including quantitative assessment of progression in the follow-up, comprehensive prediction of severity in the enrollment, and visualization of lesion distribution using percentage of infection (POI). Cao *et al.* [51] assess longitudinal progression of COVID-19 by using voxel-level deep learning-based CT segmentation of pulmonary opacities. Huang *et al.* [52] segment lung region and GGO for quantitative evaluation, which is further used for monitoring the progression of COVID-19. Qi *et al.* segment lung lesions of COVID-19 patients using a U-Net based algorithm, and extract radiomics features for predicting hospital stay [53].

In summary, image segmentation plays an important role in COVID-19 applications, i.e., in lung delineation and lesion measurement. It facilitates radiologists in accurately identification of lung infection and prompting quantitative analysis and diagnosis of COVID-19.

IV. AI-ASSISTED DIAGNOSIS

In outbreak areas, patients suspected of COVID-19 are in urgent need of diagnosis and proper treatment. Due to fast acquisition, X-ray and CT scans are widely performed to provide evidences for radiologists. However, medical images, especially chest CT, contain hundreds of slices, which takes a long time for the specialists to diagnose. Also, COVID-19 as a new disease has similar manifestations with various other types of pneumonia, which requires radiologists to accumulate many experiences for achieving a high diagnostic performance. Thus, AI-assisted diagnosis using medical images is highly desired. Table II lists the most relevant state-of-the-art studies on this direction.

*A. X-ray based Screening of COVID-19*

X-ray images are generally considered less sensitive than 3D chest CT images, despite being the typical first-line imaging modality used for patients under investigation of COVID-19. A recent study reported that X-ray shows normal in early or mild disease [71]. In particular, abnormal chest radiographs are found in 69% of the patients at the initial time of admission, and in 80% of the patients some time after during hospitalization [71].

Radiological signs include airspace opacities, ground-glass opacity (GGO), and later consolidation. Bilateral, peripheral, and lower zone predominant distributions are mostly observed (90%). Pleural effusion is rare (3%) in comparison to parenchymal abnormalities [71].

Classification of COVID-19 from other pneumonia and healthy subjects have been explored. Ghoshal *et al.* [72] propose a Bayesian Convolutional Neural network to estimate the diagnosis uncertainty in COVID-19 prediction. 70 lung X-ray images of patients with COVID-19 are obtained from an online COVID-19 dataset [73], and non-COVID-19 images are obtained from Kaggle's Chest X-Ray Images (Pneumonia). The experimental results show that Bayesian inference improves detection accuracy of the standard VGG16 model from 85.7% to 92.9%. The authors further generate saliency maps to illustrate the locations focused by the deep network, to improve the understanding of deep learning results and facilitate a more informed decision-making process.

TABLE II
RELATED STUDIES WITH MEDICAL IMAGES FOR AI-ASSISTED DIAGNOSIS OF COVID-19

| Literature | Modality | Subjects | Task | Method | Result |
|---|---|---|---|---|---|
| Ghoshal et al. [72] | X-Ray | 70 COVID-19 / Others (# of subjects not available) | Classification: COVID-19/ Others | CNN | 92.9% (Acc.) |
| Narin et al. [9] | X-Ray | 50 COVID-19 / 50 Normal | Classification: COVID-19/ Normal | ResNet50 | 98.0% (Acc.) |
| Zhang et al. [74] | X-Ray | 70 COVID-19 / 1008 Others | Classification: COVID-19/ Others | ResNet | 96.0% (Sens.) / 70.7% (Spec.) / 0.952 (AUC) |
| Wang et al. [11] | X-Ray | 45 COVID-19 / 931 Bac. Pneu. / 660 Vir. Pneu. / 1203 Normal | Classification: COVID-19/ Bac. Pneu./ Vir. Pneu./ Normal | CNN | 83.5% (Acc.) |
| Chen et al. [56] | CT | 51 COVID-19 / 55 Others | Classification: COVID-19/ Others | UNet++ | 95.2% (Acc.) / 100% (Sens.) / 93.6% (Spec.) |
| Zheng et al. [50] | CT | 313 COVID-19 / 229 Others | Classification: COVID-19/ Others | U-Net CNN | 90.7% (Sens.) / 91.1% (Spec.) / 0.959 (AUC) |
| Jin et al. [69] | CT | 496 COVID-19 / 1385 Others | Classification: COVID-19/ Others | CNN | 94.1% (Sens.) / 95.5% (Spec.) |
| Jin et al. [57] | CT | 723 COVID-19 / 413 Others | Classification: COVID-19/ Others | UNet++ CNN | 97.4% (Sens.) / 92.2% (Spec.) |
| Wang et al. [75] | CT | 44 COVID-19 / 55 Vir. Pneu. | Classification: COVID-19/ Vir. Pneu. | CNN | 82.9% (Acc.) |
| Ying et al. [70] | CT | 88 COVID-19 / 100 Bac. Pneu. / 86 Normal | Classification: COVID-19/ Bac. Pneu./ Normal | ResNet-50 | 86.0% (Acc.) |
| Xu et al. [76] | CT | 219 COVID-19 / 224 Influ.-A / 175 Normal | Classification: COVID-19/ Influ.-A/ Normal | CNN | 86.7% (Acc.) |
| Li et al. [55] | CT | 468 COVID-19 / 1551 CAP / 1445 Non-pneu. | Classification: COVID-19/ CAP/ Non-pneu. | ResNet-50 | 90.0% (Sens.) / 96.0% (Spec.) |
| Shi et al. [77] | CT | 1658 COVID-19 / 1027 CAP | Classification: COVID-19/CAP | RF | 87.9% (Acc.) / 90.7% (Sens.) / 83.3% (Spec.) |
| Tang et al. [78] | CT | 176 COVID-19 | Severity assessment | RF | 87.5% (Acc.) / 93.3% (TPR) / 74.5% (TNR) |

Bac. Pneu.: Bacterial pneumonia; Vir. Pneu.: Viral pneumonia; Influ.-A: Influenza-A; Non-pneu.: Non- pneumonia

Narin *et al.* [9] propose three different deep learning models, i.e., ResNet50, InceptionV3, and Inception-ResNetV2, to detect COVID-19 infection from X-ray images. It is worth noting that the COVID-19 dataset [73] and Kaggle's Chest X-Ray Images (Pneumonia) are also used to form the dataset in this study. Chest X-ray images of 50 COVID-19 patients and 50 normal chest X-ray images are included. The evaluation results show that the ResNet50 model achieves the highest classification performance with 98.0% accuracy, compared to 97.0% accuracy by InceptionV3 and 87% accuracy by Inception-ResNetV2.

Zhang *et al.* [74] present a ResNet based model to detect COVID-19 from X-ray images. This model has two tasks, i.e., one task for the classification between COVID-19 and non-COVID-19, and another task for anomaly detection. The anomaly detection task gives an anomaly score to optimize the COVID-19 score used for the classification. X-ray images from 70 COVID-19 patients and 1008 non-COVID-19 pneumonia patients are included from these two datasets. The sensitivity and specificity are 96.0% and 70.7%, respectively, along with an AUC of 0.952.

Also, Wang *et al.* [11] propose a deep convolutional neural network based model (COVID-Net) to detect COVID-19 cases using X-ray images. Similarly, from these two datasets, the dataset includes 5941 chest X-ray images from 1203 healthy people, 931 patients with bacterial pneumonia, 660 patients with viral pneumonia, and 45 patients with COVID-19. The COVID-Net obtains the testing accuracy of 83.5%.

In general, most current studies use X-ray images to classify between COVID-19 and other pneumonia and healthy subjects. The images are mainly from two online datasets, in which there are only 70 images from COVID-19 patients. With this limited number of COVID-19 images, it is insufficient to evaluate the robustness of the methods and also poses questions to the generalizability *with respect to* applications in other clinical centers. Also, the severity of subjects remain unknown; the future work could emphasize on early detection of COVID-19.

*B. CT-based Screening and Severity Assessment of COVID-19*

Dynamic radiological patterns in chest CT images of COVID-19 have been reported and summarized as 4 stages [79]. Briefly, 0-4 days after onset of the initial symptom is considered as the early stage. GGO could be observed subpleurally in the lower lobes unilaterally or bilaterally. The progressive stage is 5-8 days where diffuse GGO, crazy-paving pattern, and even consolidation could be found distributing in bilateral multi-lobes. In the peak stage (9-13 days), dense consolidation becomes more prevalent. When the infection becomes controlled, the absorption stage appears (usually after 14 days). Consolidation and crazy-paving pattern are gradually absorbed and only GGO is left. These radiological patterns provide important evidences for CT-based classification and severity assessment of COVID-19.

*1) Classification of COVID-19 from non-COVID-19.* There are a number of studies aiming to separate COVID-19 patients from non-COVID-19 subjects (that include common pneumonia subjects and non-pneumonia subjects). Chen *et al.* [56] employ chest CT images from 51 COVID-19 patients and 55 patients with other diseases to train a UNet++ based segmentation model. This model is responsible for segmenting COVID-19 related lesions. The final label (COVID-19 or non-COVID-19) is determined based on the segmented lesions. The evaluation results of COVID-19 classification using the proposed model are 95.2% (Accuracy), 100% (Sensitivity), and 93.6% (Specificity). In an additional dataset including 16 viral pneumonia and 11 non-pneumonia patients, the proposed model could identify all the viral pneumonia patients and 9 of non-pneumonia patients. The reading time of radiologists is shortened by 65% with the help of AI results.

A U-Net+3D CNN based model (DeCoVNet) is proposed in [50]. The U-Net is used for lung segmentation, and the segmentation result is used as the input of the 3D CNN for predicting the probability of COVID-19. Chest CT images of 540 subjects (i.e., 313 with COVID-19, and 229 without COVID-19) are used as training and testing data. The proposed model achieves a sensitivity of 90.7%, specificity of 91.1%, and AUC of 0.959.

Jin *et al.* [69] use chest CT images from 496 COVID-19 positive cases and 1385 negative cases. A 2D CNN based model is proposed to segment the lung and then identify slices of positive COVID-19 cases. Experimental results show that the proposed model achieves sensitivity of 94.1%, specificity of 95.5%, and AUC of 0.979.

In another work, Jin *et al.* [57] include chest CT images of 1136 cases (i.e., 723 COVID-19 positives, and 413 COVID-19 negatives) in the study. Their proposed model contains a UNet++ based segmentation model and a ResNet50 based classification model. The segmentation model is used to highlight lesion regions for the classification model. In the experiment, the sensitivity and specificity using the proposed UNet++ and ResNet50 combined model are 97.4% and 92.2%, respectively.

*2) Classification of COVID-19 from other pneumonia.* Given that the common pneumonia especially viral pneumonia has similar radiological appearances with COVID-19, their differentiation would be more useful in facilitating the screening process in clinical practice. Thus, a CNN model is proposed in [75] to classify between COVID-19 and typical viral pneumonia. Chest CT images from 99 patients (i.e., 44 COVID-19 and 55 typical viral pneumonia) are used. Slices derived from each 3D CT image are used as the input of the proposed CNN. The testing dataset shows a total accuracy of 73.1%, along with a specificity of 67.0% and a sensitivity of 74.0%.

Ying *et al.* [70] proposea deep learning-based CT diagnosis system (called DeepPneumonia, using ResNet50) to detect patients with COVID-19 from bacteria pneumonia patients and healthy people. Chest CT images from 88 patients with COVID-19, 101 patients with bacterial pneumonia, and 86 healthy persons are used as training and testing data. Slices of complete lungs are derived from chest CT images, which are the inputs of DeepPneumonia. The model achieves promising results with an accuracy of 86.0% for pneumonia classification (COVID-19 or bacterial pneumonia), and an accuracy of 94.0% for pneumonia diagnosis (COVID-19 or healthy).

Xu *et al.* [76] use chest CT images from 219 patients with COVID-19, 224 patients with Influenza-A, and 175 healthy persons. A deep learning model based on V-Net is first used to segment out the candidate infection regions. The patches of infection regions are sent to the Resnet-18 network together with features of relative infection distance from edge, and the output is one of these three groups. The model achieves an overall accuracy of 86.7%.

Li *et al.* [55] used a large chest CT dataset, which contains 4356 chest CT images (i.e., 1296 COVID-19, 1735 community-acquired pneumonia, and 1325 non-pneumonia) from 3322 patients. A ResNet50 model (COVNet) is used on 2D slices with shared weights and combined with max-pooling to discriminate COVID-19 from community-acquired pneumonia (CAP) and non-pneumonia. Results show a sensitivity of 90%, specificity of 96%, and AUC of 0.96 in identifying COVID-19.

Shi *et al.* [77] employed chest CT images of 2685 patients, of which 1658 patients are of COVID-19 and 1027 patients are of common pneumonia. In the preprocessing stage, a VB-Net [58] is adopted to segment the image into the left/right lung, 5 lung lobes, and 18 pulmonary segments. A number of hand-crafted features are calculated and used to train a random forest model. Experimental results show a sensitivity of 90.7%, specificity of 83.3%, and accuracy of 87.9%. Also, testing results are grouped based on infection sizes, showing that patients with small infections have low sensitivity to be identified.

*3) Severity assessment of COVID-19.* Besides early screening, the study of severity assessment is also important for treatment planning. Tang *et al.* [78] proposed an RF-based model for COVID-19 severity assessment (non-severe or severe). Chest CT images of 176 patients with conformed COVID-19 is used. A deep learning method VB-Net [77] is adopted to divide the lung into anatomical sub-regions (e.g., lobes and segments), based on which infection volumes and ratios of each anatomical sub-region are calculated and used as quantitative features to train a RF model. Results show a true positive rate of 93.3%, true negative rate of 74.5%, and accuracy of 87.5%.

In summary, a variety of studies have been proposed for CT-based COVID-19 diagnosis with generally promising results. In the next step, the research on screening of COVID-19 could facilitate early detection to help with the diagnosis uncertainty of radiologists. Also, the prediction of severity is of great importance that could help the estimation of the ICU event or clinical decision of treatment planning, which warrants more investigation.

## V. AI IN FOLLOW-UP STUDIES

With the goal of evaluating the patient's response and investigating their potential problems after clinical treatment, the follow-up step plays a significant role in COVID-19 treatment. Regarding the long incubation period of COVID-19 and its popular infectivity, to design the procedure of AI-empowered follow-up for COVID-19 is challenging.

As most of the current works focus on the pre-diagnosis of COVID-19, we notice that the works for studying the follow-up for COVID-19 are still very limited. There are only few attempts according to our knowledge. For example, the researchers of Shanghai United Imaging Intelligence (UII) attempt to use the machine learning-based method and visualization techniques to demonstrate the change of the volume size, density, and other clinical related factors in the infection regions of the patient. After that, the clinical report is automatically generated to reflect these changes as a data-driven guidance for clinical specialists to determine the following procedure (Fig. 2). In addition, the team from Perception Vision Company (PVmed) provided another follow-up solution for COVID-19. They tried to build a contrastive model to reflect the change of different CT images of the same patient, by aligning the infection regions and observing the changing trend of these quantitative values. Several other companies and institutes are also developing the follow-up function in their software platforms currently. Subsidiarily, Huang *et al.* [52] collect and analyze 126 patients by calculating the CT lung opacification percentage. They find the quantification of lung involvement could be employed to reflect the disease progression of COVID-19, which is helpful for the follow-up study.

It is worth noting that clinical specialists are taking their efforts to the diagnosis and treatment of COVID-19. Thus, the works for studying the follow-up of COVID-19 are still in the early stage and remain an open issue. We believe the previous techniques and work developed in segmentation, diagnosis, quantification, and assessment could be used to guide the development of AI-empowered follow-up study for COVID-19.

## VI. PUBLIC IMAGING DATASETS FOR COVID-19

Data collection is the first step to develop machine learning methods for COVID-19 applications. Although there exist large public CT or X-ray datasets for lung diseases, both X-ray and CT scans for COVID-19 applications are not widely available at present, which greatly hinders the research and development of AI methods. Recently, several works on COVID-19 data collection have been reported.

Cohen *et al.* [73] creates COVID-19 Image Data Collection by assembling medical images from websites and publications, and it currently contains 123 frontal view X-rays. The COVID-CT dataset [80] includes 288 CT slices for COVID-19 confirmed cases thus far. It is collected from over 700 preprinted literature on COVID-19 from medRxiv and bioRxiv. The Coronacases Initiative also shares confirmed cases of COVID-19 on the website (https://coronacases.org). Currently, it includes 3D CT images of 10 confirmed COVID-19 cases. Also, the COVID-19 CT segmentation dataset (http://medicalsegmentation.com/covid19/) contains 100 axial CT slices from 60 patients with manual segmentations, in the form of JPG images. It is worth noting that the current public datasets still have a very limited number of images for training and testing of AI algorithms, and the quality of datasets is not sufficient .

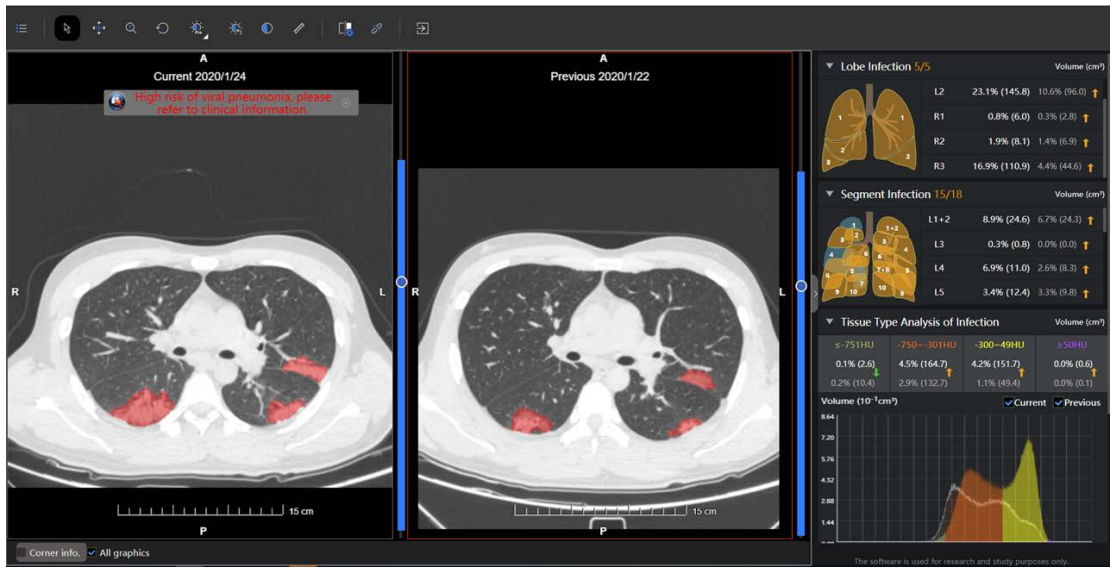

Fig. 2. The follow-up measurement for a COVID-19 patient.

## VII. DISCUSSION AND FUTURE WORK

AI has been successfully applied to the entire pipeline of the imaging-based diagnosis of COVID-19. However, there are still many works to be conducted in the future.

As mentioned, AI-empowered image acquisition workflows have proven to make the scanning procedure *not only* more efficient, *but also* effective in protecting medical staffs from COVID-19 infection. Looking ahead, it is expected that more AI-empowered applications will be integrated into the image acquisition workflow, to facilitate better scan quality and reduced radiation dosage consumed by patients. For example, more precise AI-based automated ISO-centering and scan range determination are required to ensure optimal image quality. Moreover, X-ray exposure parameters can be automatically calculated and optimized with AI inferred body region thickness of the patient, ensuring that just the right amount of radiation is used during the scan, which is particularly important for low-dose imaging.

Medical images usually show negative radiological signs in the early stage of the disease, and thus the study of this stage is important to assist with the clinical diagnosis uncertainty. Meanwhile, many current AI studies for segmentation and diagnosis are based on small samples, which may lead to the overfitting of results. To make the results clinically useful, the quality and number of data need to be further improved. Therefore, more datasets should be built to include the clinically collected X-ray and CT images.

Deep learning has become the dominant approach in fighting against COVID-19. However, the imaging data in COVID-19 applications may have incomplete, inexact and inaccurate labels, which provides a challenge for training an accurate segmentation and diagnostic network. In this way, weakly supervised deep learning methods could be leveraged. Further, manually labeling imaging data is expensive and time-consuming, which also encourages the investigation of self-supervised deep learning and transfer-deep-learning methods. Multi-center studies on COVID-19 should also be promoted.

Follow-up is critical in diagnosing COVID-19 and evaluating treatment. Although there are still limited studies, we believe that the methods from other related studies could be borrowed. 1) In the prognosis of other pneumonia diseases, machine learning-based methodology could inspire the follow-up study of COVID-19 [81-84]. 2) The follow-up inside and outside of hospitals could be combined as a long period tracking for the COVID patients. 3) Multidisciplinary integration, i.e., medical imaging [85], natural language processing [86], and oncology and fusion [86], could benefit the overall follow-up procedure of measurement for COVID-19.

## VIII. CONCLUSION

The COVID-19 is a disease that has spread all over the world. Intelligent medical imaging has played an important role in fighting against COVID-19. This paper discusses how AI provides safe, accurate and efficient imaging solutions in COVID-19 applications. The intelligent imaging platforms, clinical diagnosis, and pioneering research are reviewed in detail, which covers the entire pipeline of AI-empowered imaging applications in COVID-19. Two imaging modalities, i.e., X-ray and CT, are used to demonstrates the effectiveness of AI-empowered medical imaging for COVID-19.

It is worth noting that imaging only provides partial information about patients with COVID-19. It is important to combine imaging data with clinical manifestations and laboratory examination results to help the screening, detection and diagnosis of COVID-19. In this case, we believe AI will demonstrate its capability in fusing information from these multi-source data, for performing accurate and efficient diagnosis, analysis and follow-up.